\documentclass[preprint,12pt]{elsarticle}

\usepackage{graphicx,subfigure,amsmath,amssymb,epstopdf,float,url}

\journal{Physica A}

\begin{document}

\begin{frontmatter}

\title{Efficient construction of threshold networks of stock markets}

\author[1]{Xin-Jian Xu}
\author[1]{Kuo Wang}
\author[2]{Liucun Zhu}
\author[3]{Li-Jie Zhang\corref{corresponding}}
\ead{lijzhang@shu.edu.cn}
\cortext[corresponding]{Corresponding author}

\address[1]{Department of Mathematics, Shanghai University, Shanghai 200444, China}
\address[2]{School of Life Sciences, Shanghai University, Shanghai 200444, China}
\address[3]{Department of Physics, Shanghai University, Shanghai 200444, China}

\begin{abstract}
Although the threshold network is one of the most used tools to characterize the underlying structure of a stock market, the identification of the optimal threshold to construct a reliable stock network remains challenging. In this paper, the concept of dynamic consistence between the threshold network and the stock market is proposed. The optimal threshold is estimated by maximizing the consistence function. The application of this procedure to stocks belonging to Standard \& Pool's 500 Index from January 2006 to December 2011 yields the threshold value 0.28. In analyzing topological characteristics of the generated network, three globally financial crises can be distinguished well from the evolutionary perspective.
\end{abstract}


\end{frontmatter}

\section{Introduction}

Stock markets are well-defined complex systems, consisting of interacting stocks and instruments~\cite{ms00book}. Studying global complexity of the invisible market is a significant issue, which plays an important role in practical applications such as fluctuation prediction and asset allocation~\cite{clm97book}. To characterize statistical structure of multiple time series of stock prices, random matrix theory (RMT)~\cite{plerou99prl} was applied to study the eigenvalue distribution of the correlation coefficient matrix of time series. There are stylized facts of stock markets unveiled by the RMT~\cite{plerou02pre,eom09phya,namaki11phya,song11pre,meng14sr,jiang14sr,dai16ee,wang17chaos,han17fnl}, for example, a market contains many business sectors (communities of stocks sharing common economic properties) with hierarchial organization. However, the RMT could not draw interactions well among these sectors.

On the other hand, complex network theory (CNT)~\cite{newman10book} was used to visualize and understand core information of stock markets. To transform a stock market into a network, the first step is to measure interactions among stocks. There are several measurements including linear correlation~\cite{mantegna99epjb}, phase synchronization~\cite{ss11conf} and mutual information~\cite{fiedor14pre,yang14mplb}. Mantegna was the first~\cite{mantegna99epjb} to construct stock networks based on the minimal spanning tree (MST)~\cite{bm08book}.
Further refinement of this method, named the planar maximally filtered graph (PMFG), was done by Tumminello et al~\cite{tumminello05pnas}. These two approaches were widely adopted in the analysis of financial markets such as New York Stock Exchange~\cite{mantegna99epjb,tumminello07ijbc,tumminello07epjb,wang17jeic}, Brazilian Stock Market~\cite{tabak10phya}, German Stock Exchange~\cite{wilinski13phy}, Shanghai Stock Market~\cite{yang13mplb}, and South African Stock Market~\cite{mg16phya}.

Due to topological limit, however, some significant edges of high similarity will be excluded by the MST and the PMFG. To overcome this shortcoming, Boginski et al. proposed a threshold to discard all correlations less than it and constructed a threshold network (TN)~\cite{boginski05csda}, which triggered a series of studies~\cite{huang09phya,liu11qf,nobi14phya,chu17phya,xia18phya}. Among research, a fundamental question is to identify accurate threshold to construct reliable stock networks. Recent work suggested statistical text method to find the solution~\cite{koldanov13csda,ha15jkps,ak15conf,valeriy16amai,xu17sr}, but this methodology usually doesn't account for multiple hypothesis testing corrections~\cite{mg15prx}.

In this paper, we aim to introduce a novel method to estimate the optimal threshold to construct a stock network which can capture the essential features of a stock market. Although the underlying structure of the market is a scientific black-box, there are certain observable quantities reflecting it. As to the generated network, there are also several topological parameters characterizing it. Given a time horizon, both financial quantities and network parameters evolve simultaneously. The change in an idealized network should be consistent with that in the real market. Motivated by this, we introduce a consistent function between them, and estimate the optimal threshold conditional on maximal consistence. We apply this framework to stocks belonging to Standard \& Pool's (S\&P) 500 Index from January 2006 to December 2011 and yield the optimal value of the threshold. The structural statistics of the corresponding network reveals financial crises from the evolutionary perspective.

\section{Model}

Denoting with $p_i(t)$ the price of stock $i$ at time $t$, one can calculate the logarithmic price return of $i$ over a time interval $\tau$ by
\begin{equation}\label{return}
r_i(t)=\ln p_i(t)-\ln p_i(t-\tau).
\end{equation}
Then, the cross-correlation coefficient between stocks $i$ and $j$ is defined by
\begin{equation}\label{correlate}
w_{ij}=\frac{\langle r_i(t) r_j(t)\rangle - \langle r_i(t)\rangle \langle r_j(t)\rangle }{\sigma_i \sigma_j},
\end{equation}
where $\langle \cdots \rangle$ represents temporal average over $\tau$ and $\sigma_i$ is the standard deviation of $r_i(t)$. The ensemble of $w_{ij}$ forms the correlation matrix $W$ of a stock market. To construct a TN of stocks, a certain value of the threshold $\theta$ is specified. For each pair of stocks $i$ and $j$, an edge is created between them if $w_{ij} \geq \theta$. This process is repeated throughout all the elements of the matrix, and finally the TN is generated.

To find the optimal threshold $\theta$, we assume that the TN is a statistical variable derived from a parametric model
\begin{equation}\label{modelpara}
N \sim F(W, \theta),
\end{equation}
where $N$ represents the TN, $W$ represents the specified market, and $\theta$ is the parameter dominating the TN. Following the idea of parameter estimation which makes the artificial network consistent with observable quantities, we introduce a consistent function $G(N, W)$ between them. The parameter $\theta$ can be estimated by maximizing $G$:
\begin{equation}
\hat{\theta}=\mathop{\arg\max}_{\theta\in\Theta}\\G(N, W).
\end{equation}
Since it is hard to compare any network parameters to the correlation coefficient matrix directly, we shall consider dynamic consistence by rewriting the consistent function as
\begin{equation}\label{functdc}
G(D(N_t,N_{t+\tau}), D(W_t,W_{t+\tau})),
\end{equation}
where $D(N_t,N_{t+\tau})$ and $D(W_t,W_{t+\tau})$ represent the differences between any two successive networks and matrices, respectively. Although it is easy to calculate $D(W_t,W_{t+\tau})$ according to the matrix theory, e.g., \begin{equation}\label{diffmatrix}
D(W_t,W_{t+\tau})=||W_t-W_{t+\tau}||_2,
\end{equation}
there are very few methods to quantify $D(N_t,N_{t+\tau})$. In the present work, we employ ideas proposed by Schieber et al~\cite{schieber2017nc} to compute $D(N_t,N_{t+\tau})$, which is defined by
\begin{equation}\label{diffnet}
\begin{aligned}
D(N_t,N_{t+\tau}) & = \alpha\sqrt{\frac{\mathit{J}(P_l(N_t),P_l(N_{t+\tau}))}{\log{2}}}\\ &+\beta|\sqrt{H_l(N_t)}-\sqrt{H_l(N_{t+\tau})}|\\
&+\frac{\gamma}{2}\left[\sqrt{\frac{\mathit{J}(P_{\alpha}(N_t),P_{\alpha}(N_{t+\tau}))}{\log{2}}}+\sqrt{\frac{\mathit{J}(P_{\alpha}(N'_t),P_{\alpha}(N'_{t+\tau}))}{\log{2}}}\right].
\end{aligned}
\end{equation}
The first term on the right-hand side of Eq.~(\ref{diffnet}) considers the difference in vertex distances. $P_l(N_t)$ is vertex distance distribution of network $N_t$. The second term captures the difference in vertex dispersion. $H_l(N_t)$ characterizes the network heterogeneity in terms of the shortest path length, defined by
\begin{equation}\label{lengthheter}
H_l(N_t) = \frac{\mathit{J}(P_l(1),\cdots,P_l(n))}{\log{(\lambda+1)}},
\end{equation}
where $P_l(i)$ $(i=1,\cdots,n)$ is the distance distribution of vertex $i$ and $\lambda$ is the diameter of the network.
$\mathit{J}(P_l(1),\cdots,P_l(n))$ is Jensen-Shannon divergence, defined by
\begin{equation}\label{defjsd}
\mathit{J}(P_l(1),\cdots,P_l(n))=S\left(\frac{\sum_i P_l(i)}{n}\right)-\frac{\sum_i S(P_l(i))}{n},
\end{equation}
where $S$ is Shannon entropy. The third term analyzes the difference in vertex centrality. $P_{\alpha}(N_t)$ is vertex centrality distribution of network $N_t$ and $N'_t$ is the complement of $N_t$. $\alpha$, $\beta$, and $\gamma$ are
arbitrary weights of the terms with $\alpha+\beta+\gamma=1$. Following Ref.~\cite{schieber2017nc} we
selected the following weights $\alpha=\beta=0.45$ and $\gamma=0.1$ in the present work.

Given the time interval $\tau$, one can calculate the dynamic consistence (\ref{functdc}) based on Eqs.~(\ref{diffmatrix}) and (\ref{diffnet}). However, the terms on the right-hand side of Eq.~(\ref{diffnet}) are so complicated that it is hard to obtain any analytical solution to the consistent function. As an alternative, we introduce a numerical method. From the definition of the correlation coefficient (Eq.~(\ref{correlate})), it follows that $\theta \in [-1, 1]$. Then, one can sample $\theta$ uniformly from $[-1,1]$ and obtain an ascending sequence $\{\theta_1, \theta_2,\cdots,\theta_n\}$ with $\theta_{1}=-1$ and $\theta_n=1$. At each threshold $\theta_{i}$, a correlation matrix and TN can be generated using the moving-window method~\cite{onnela03pre}. So one has two sequences: $\{W_{1},W_{2},\cdots,W_{m}\}$ and $\{N_{1},N_{2},\cdots,N_{m}\}$. According to Eqs.~(\ref{diffmatrix}) and (\ref{diffnet}), it is easy to obtain two differing sequences: $\{D(W_1,W_2),D(W_2,W_3),\cdots,D(W_{m-1},W_m)\}$ and $\{D(N_1,N_2),D(N_2,N_3),\cdots,D(N_{m-1},N_m)\}$. For any $\theta_{i}$, one can measure the consistence between the changes in the correlation matrix and the network by Pearson Correlation Coefficients,
\begin{equation}\label{defpcc}
G_{\theta_{i}}=\frac{\langle D_{W}D_{N}\rangle-\langle D_{W}\rangle\langle D_{N}\rangle}{\sigma_{D_{W}}\sigma_{D_{N}}},
\end{equation}
where $\langle D_{W}\rangle$ and $\langle D_{N}\rangle$ are the means of changes in the matrix and the network, respectively. $\sigma_{D_W}$ and $\sigma_{D_N}$ are corresponding standard deviations. Finally, the optimal threshold can be estimated from the numerical way
\begin{equation}
\hat{\theta}=\arg\max_{\theta_{i}}\{G_{\theta_{i}}\}.
\end{equation}

\section{Application to S\&P $500$ Stocks}

\begin{figure}[h]
\includegraphics[width=\columnwidth]{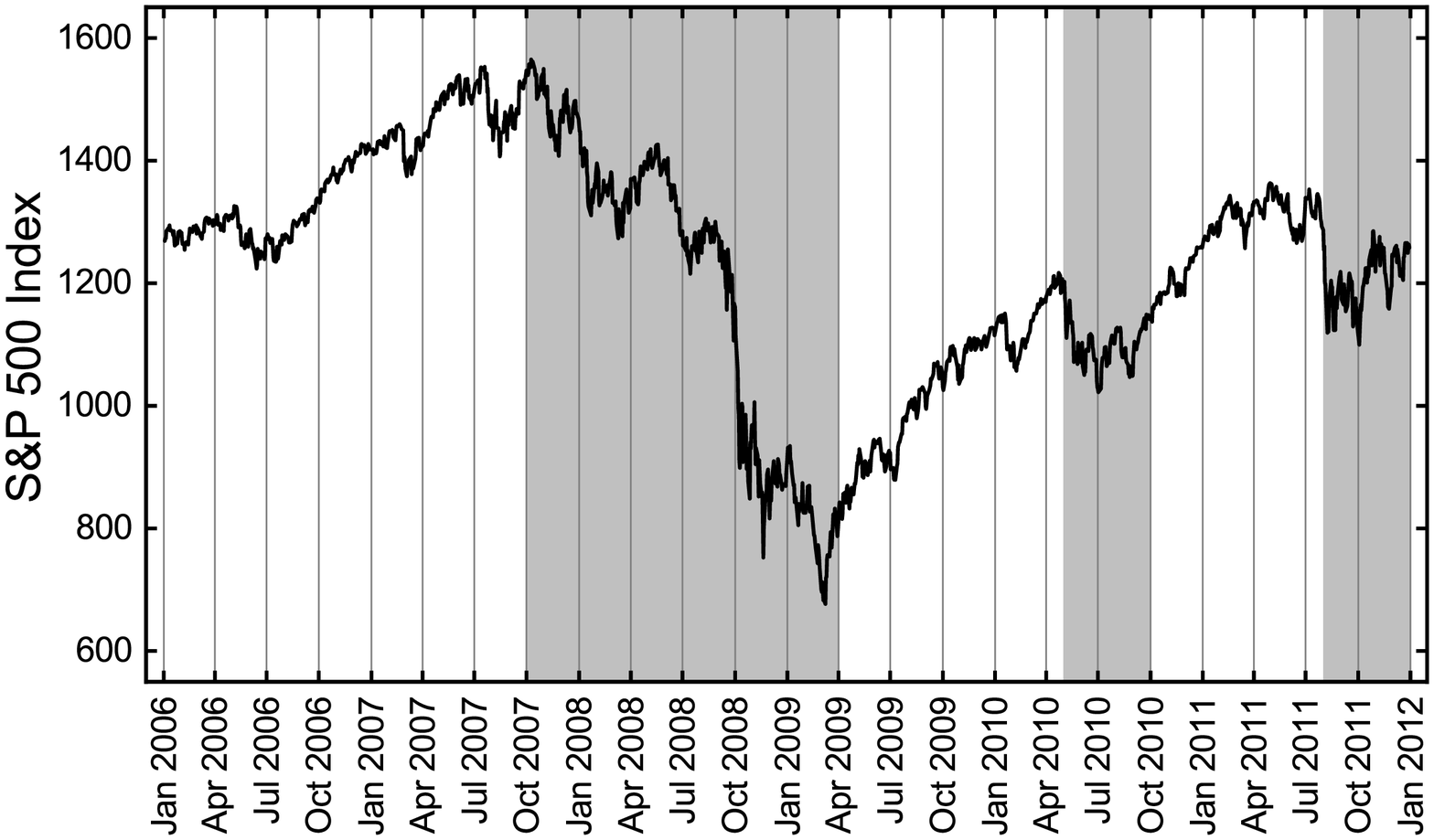}
\caption{Time series of S\&P 500 Index from January 2006 to December 2011. The gray bars correspond to three financial crises: the 2008 subprime crisis, the 2010 Greek debt crisis, and the 2011 European sovereign debt crisis.}\label{fig1}
\end{figure}

To test its validity, we apply the above method to a set of 445 stocks belonging to S\&P 500 Index.
The data are daily records and the investigated period ranges from January 2006 to December 2011 consisting of 1511 observations. During this period, there were three main financial crises of S\&P 500 Market: the 2008 subprime crisis, the 2010 Greek debt crisis, and the 2011 European sovereign debt crisis. Figure~\ref{fig1} depicts the trend of S\&P 500 Index where gray intervals correspond to three financial crises, in contrast to the periods of business as usual, bull and bear runs. Thus, different market states are contained in the whole scale.

\begin{figure}[h]
\includegraphics[width=\columnwidth]{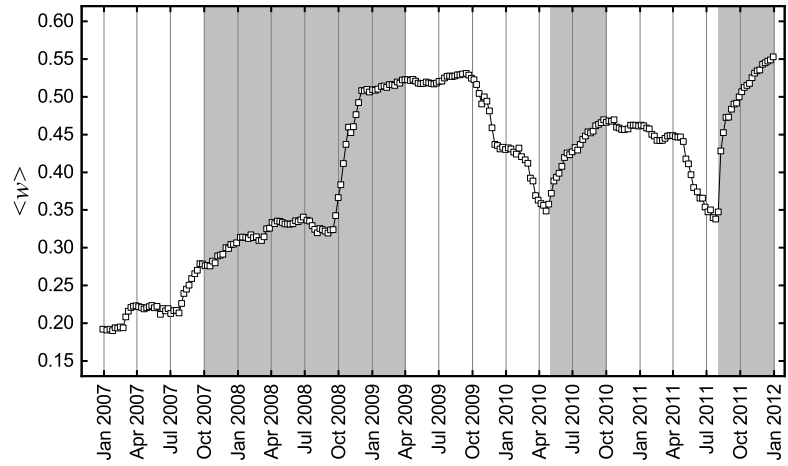}
\caption{Time series of the average correlation coefficient $\langle w\rangle$ from January 2006 to December 2011.}\label{fig2}
\end{figure}

To analyze dynamic properties of S\&P 500 Index, we divide the whole period by the moving-window method. Following Majapa et al.~\cite{mg16phya}, Alkan et al.~\cite{ak15conf}, and Onnela et al.~\cite{onnela03pre}, we set the width of each window as $t = 250$ days and the moving step as $\tau = 5$ days. With these choices, the overall number of windows is $253$. Figure~\ref{fig2} shows the evolution of the average correlation coefficient $\langle w\rangle$ of the investigated period, defined by
\begin{equation}\label{eqacc}
\langle w\rangle=\frac{\sum_{i,j}{w_{i,j}}}{N(N-1)} \quad (i\neq j).
\end{equation}
It turns out that there exists significant difference between financial crises and usual periods. At the begging of a crisis, $\langle w\rangle$ will increase rapidly. After reaching a high plateau, it will stay at the high value and even after the crisis until decreasing to the low value. Among three crises, $\langle w\rangle$ in the 2010 Greek debt crisis is lower than the other two crises which have wider and deeper impacts.

\begin{figure}[h]
\includegraphics[width=\columnwidth]{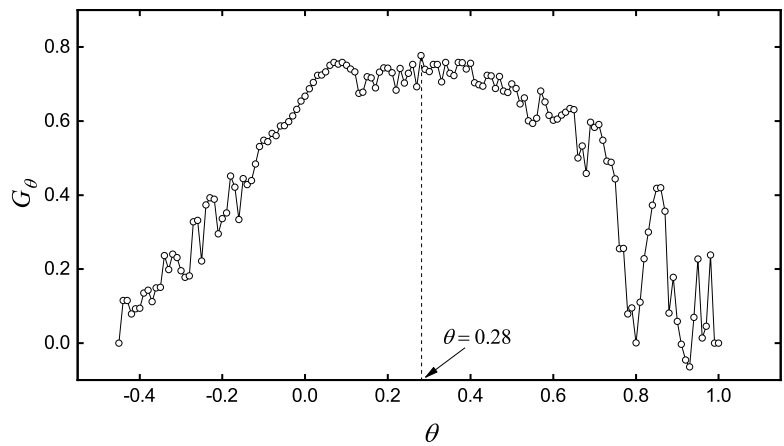}
\caption{Dynamic consistence $G_{\theta}$ as a function of $\theta$.}\label{fig3}
\end{figure}

To estimate the optimal threshold $\theta$, we calculate all pairs of stocks, resulting in the minimal correlation $-0.44$ and the maximal correlation $0.99$. By setting $\theta_{1} = -0.45$ and $\Delta\theta=0.01$, we have the threshold sequence $\{-0.45, \cdots, 1\}$ with $146$ discrete values. At each value $\theta_i$ of the threshold, we construct a corresponding network along the investigated window. After that, we calculate the consistent function $G_{\theta_i}$ between the network and the correlation coefficient according to~(\ref{functdc}). In case of $\tau = 5$, the difference in two successive matrices and the dissimilarity between two neighboring networks are relative small, but we compute the relative dynamic consistence $G_{\theta}$ between them instead of absolute values. As demonstrated in Fig.~\ref{fig3}, $G_{\theta}$ reaches the maximum at $\theta=0.28$, resulting in the optimal value of the threshold. In the following, we shall use $\theta=0.28$ to construct the optimal stock network for each time window and analyze its topological parameters.

\begin{figure}[h]
\includegraphics[width=\columnwidth]{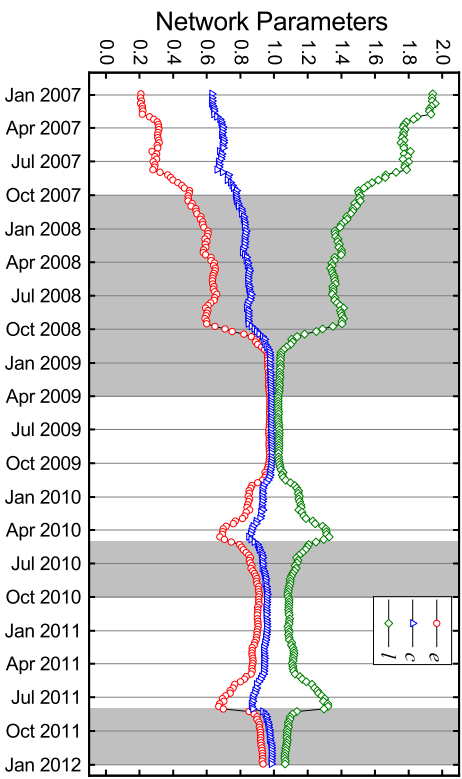}
\caption{(Color online) Evolution of the edge density, the clustering coefficient, and the shortest path length of the optimal network.}\label{fig4}
\end{figure}

First, we investigate the edge density, the average clustering coefficient, and the average shortest path length of generated networks. The edge density reflects the proportion of stock interactions remained in the network, defined by
\begin{equation}\label{defedge}
e=\frac{\sum_{i,j}{I(w_{i,j}>\theta)}}{n(n-1)} \quad (i\neq j),
\end{equation}
where $I(\cdot)$ is the indicator function.
The clustering coefficient reveals the clustering tendency of vertices in the network. For the stock network, the clustering coefficient of stock $i$ is defined as the ratio of the actual number $m_{i}$ of edges over maximal number of possible interactions among its neighbors, $c_{i}=2m_{i}/[k_{i}(k_{i}-1)]$, where $k_{i}$ is the degree of stock $i$. Then the average clustering coefficient over all stocks will be
\begin{equation}\label{defcluster}
c=\frac{\sum_i{c_i}}{n}.
\end{equation}
The shortest path length between two vertices $l_{i,j}$ is defined as the minimum number of intermediate vertices that must be traversed to go from vertex to vertex. The average shortest path length is the average of $l_{i,j}$ over all the possible pairs of vertices in the network,
\begin{equation}\label{deflength}
l=\frac{\sum_{i,j}{l_{i,j}}}{n(n-1)} \quad (i\neq j).
\end{equation}
In Fig.~\ref{fig4}, we present temporal behaviors of the edge density $e$ (open circles), the average clustering coefficient $c$ (open triangles), and the average shortest path length $l$ (open diamonds). Both $e$ and $c$ display the same trend with $\langle w\rangle$, while $l$ evolves in the opposite way, since large $e$ and $c$ shorten the distance among vertices. Moreover, there exists substantial difference between financial crises and usual periods. During three crises, $e$ and $c$ increase from lower values to higher values. Especially for the 2008 subprime crisis, there exists a turning point around September 2008.  Before it, $e$ and $c$ increase at a relatively lower speed. After it, they come up with a sharply rise and then remain in the high level. This transition corresponds to the bankruptcy of Fannie Mae, Freddie MacIn and Lehman Brothers investment bank, which makes the crisis spread from the United States to the world.

\begin{figure}[h]
\includegraphics[width=\columnwidth]{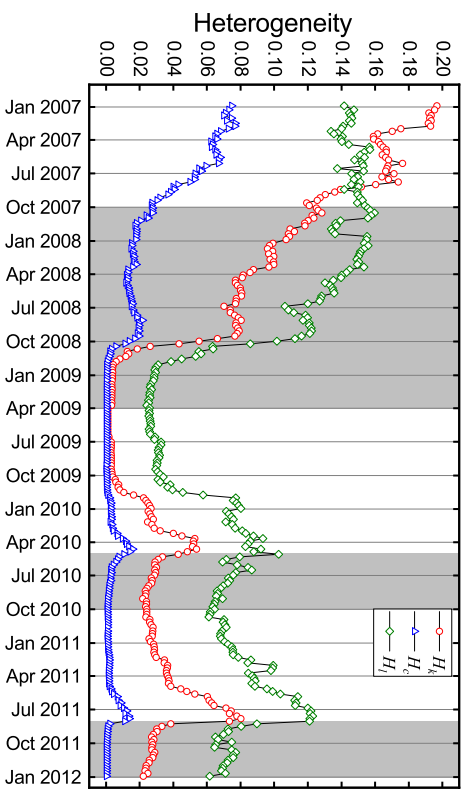}
\caption{(Color online) Evolution of the network heterogeneity in terms of the degree, the clustering coefficient, and the shortest path length, respectively.}\label{fig5}
\end{figure}

Second, we probe the network heterogeneity in terms of the degree, the clustering coefficient, and the shortest path length, which can reflect structural deviation from regularity. Following the idea of Estrada~\cite{estrada10pre}, the network heterogeneity from the perspective of vertex degrees is defined by
\begin{equation}\label{degreeheter}
H_k=\frac{\sum_{i,j\in m}{({k_{i}}^{-\frac{1}{2}}-{k_{j}}^{-\frac{1}{2}})^2}}{n-2\sqrt{n-1}},
\end{equation}
where $m$ represents the edge set of the network. The heterogeneity index $H_k$ is zero for any regular network and one for the star graph, i.e., $0 \le H_k \le 1$. Similarly, the network heterogeneity from the perspective of the clustering coefficient can be written as
\begin{equation}\label{clusterheter}
H_c=\frac{\sum_{i,j\in m}{({c_{i}}^{-\frac{1}{2}}-{c_{j}}^{-\frac{1}{2}})^2}}{n-2\sqrt{n-1}}.
\end{equation}
Based on Eqs.~(\ref{lengthheter}), (\ref{degreeheter}), and (\ref{clusterheter}), one can compute the heterogeneity of stock networks. As shown in Fig.~\ref{fig5}, all the three heterogeneity indices evolve with the same trend.
Especially during three crises, the indices decrease with time, indicating that the stock network evolves from the star-like structure to the regular one. After each crisis, however, the stock network will evolve with the opposite tendency.

\begin{figure}[h]
\includegraphics[width=\columnwidth]{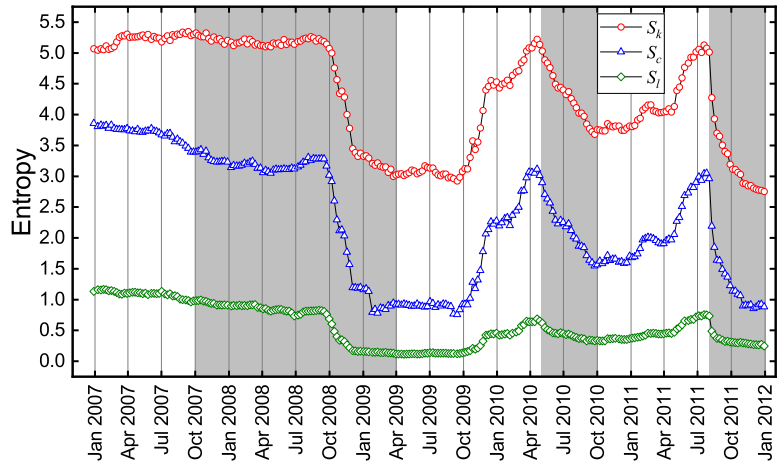}
\caption{(Color online) Evolution of the network entropy from the aspects of the degree, the clustering coefficient, and the shortest path length, respectively.}\label{fig6}
\end{figure}

Finally, we explore the network entropy in terms of the degree, the clustering coefficient, and the shortest path length, which can reflect structural diversity from regularity~\cite{dm05phya}.
Following the concept of Shannon entropy, the network entropy from the aspect of vertex degrees is defined by
\begin{equation}\label{degreeentropy}
S_k =-\sum_{k=k_\text{min}}^{k_\text{max}}p_k\text{log}p_k.
\end{equation}
Similarly, the network entropies from the aspects of the clustering coefficient and the shortest path length can be written as
\begin{equation}
S_c =-\sum_{c=c_\text{min}}^{c_\text{max}}p_{c}\text{log}p_{c}
\end{equation}
and
\begin{equation}
S_l =-\sum_{l=l_\text{min}}^{l_\text{max}}p_l\text{log}p_l,
\end{equation}
respectively. $p_{c}$ represents the distribution of the clustering coefficient of vertices. Fig.~\ref{fig6} shows temporal behavior of the network entropy in terms of the degree, the clustering coefficient, and the shortest path length, respectively. Again, one notices the decrease of three entropies during each crisis, which implies the trend of lower diversity contrary to usual periods of higher diversity. Comparing Figs.~\ref{fig5} and \ref{fig6}, we conclude that network parameters as a function of vertex degrees perform well in manifesting financial crises. However, it does not imply that these metrics can be used to predict financial crises based on the present model.

\section{Conclusion}

Most studies of stock markets by TNs are usually limited by the assertion of the threshold. In this paper, we have proposed an efficient method to estimate the optimal threshold for constructing stock networks. Suppose that both observable financial quantities and artificial network parameters are reflections of the stock market, the evolution of the two aspects should be consistent. Based on this assumption, we introduced a function of dynamic consistence (\ref{functdc}) and used the idea of parameter estimation to find the optimal threshold conditional on maximal consistence.

To test the validity of the above approach, we collected real data of stocks belonging to S\&P $500$ Index from January $2006$ to December $2011$ and divided the whole period by the moving-window method. For any two successive windows, we calculated the difference in the observable correlation matrices by Eq.~(\ref{diffmatrix}) and the difference in the artificial stock networks by Eq.~(\ref{diffnet}), respectively, so that we
can solve the consistent function (\ref{functdc}). Applying this procedure to all the successive windows, we found that the consistent function reaches the maximum at $\theta=0.28$, indicating the optimal value of the threshold for generating stock networks. With this optimal threshold, we constructed reliable stock networks. In contrast to most studies paying attention to the static structure, we focused on dynamic features of the generated network. Through exploring the edge density, the clustering coefficient, the shortest path length, the heterogeneity and the entropy, we distinguished prominently three financial crises from usual periods. Moreover, we found that the $2008$ subprime crisis exhibits different patterns from the other two crises. Therefore, the present study provides an efficient approach for building TNs.

\section*{Acknowledgments}
This work was partly supported by Natural Science Foundation of China under Grant No. 11331009 and Science and Technology Commission of Shanghai Municipality under Grant No. 17ZR1445100.

\newpage

\end{document}